\documentclass[%
reprint,
superscriptaddress,
 amsmath,amssymb,
 aps,
 pra,
]{revtex4-1}

\usepackage{graphicx}
\usepackage{dcolumn}
\usepackage{bm}
\usepackage[mathlines]{lineno}

\begin{document}


\title{Spatial confinement of muonium atoms}
\author{K.~S.~Khaw}
\email[]{Present address: Department of Physics, University of Washington, Seattle, WA 98195, U.S.A.}

\affiliation{Institute for Particle Physics, ETH Zurich, 8093 Hoenggerberg, Switzerland}
\author{A.~Antognini}
\affiliation{Institute for Particle Physics, ETH Zurich, 8093 Hoenggerberg, Switzerland}
\affiliation{Laboratory for Particle Physics, Paul Scherrer Institute, 5232 Villigen PSI, Switzerland}
\author{T.~Prokscha}
\affiliation{Laboratory for Muon Spin Spectroscopy, Paul Scherrer Institute, 5232 Villigen PSI, Switzerland}
\author{K.~Kirch}
\affiliation{Institute for Particle Physics, ETH Zurich, 8093 Hoenggerberg, Switzerland}
\affiliation{Laboratory for Particle Physics, Paul Scherrer Institute, 5232 Villigen PSI, Switzerland}
\author{L.~Liszkay}
\affiliation{IRFU, CEA, University Paris-Saclay F-91191 Gif-sur-Yvette Cedex, France}
\author{Z.~Salman}
\affiliation{Laboratory for Muon Spin Spectroscopy, Paul Scherrer Institute, 5232 Villigen PSI, Switzerland}
\author{P.~Crivelli}
\email[]{crivelli@phys.ethz.ch}
\affiliation{Institute for Particle Physics, ETH Zurich, 8093 Hoenggerberg, Switzerland}

\date{\today}
\begin{abstract}
We report the achievement of spatial confinement of muonium atoms (the bound state of a positive muon and an electron). Muonium emitted into vacuum from mesoporous silica reflects between two SiO$_2$ confining surfaces separated by 1 mm. From the data, one can extract that the reflection probability on the confining surfaces kept at 100 K is about 90\% and the reflection process is well described by a cosine law. This technique enables new experiments with this exotic atomic system and is a very important step towards a measurement of the 1S-2S transition frequency using continuous wave laser spectroscopy. 
\end{abstract}

\pacs{Valid PACS appear here}
\maketitle


\section{\label{sec:level1}Introduction}
Since its discovery by Hughes et al. \cite{Hughes1970},  Muonium (M), the bound state of an electron and a positive muon, has been a subject of extensive research. Being made of two leptons, muonium is an ideal system to study bound state QED free of finite size effects and in which hadronic corrections are strongly suppressed compared to hydrogen \cite{savely}. Study of its properties led to the determination of the fine structure constant (now best known by the electron g-2 measurement \cite{Gabrielse}),  the mass and magnetic moment of the muon and the best verification of charge equality in the first two generations of particles  \cite{LiuHFS,Meyer2000}.  Muonium has also been used for searches of new physics  \cite{Willmann} and found its application in materials science \cite{reviewMuAppl}. 
Due to the limited lifetime of the muon (2.2 $\mu$s), muonium is unstable. It is produced by combining $\mu^+$ from a beam with an electron of a target material. New experiments with this atomic system would benefit enormously from more intense and brighter sources and more efficient converters of the primary beam into muonium  \cite{Jungmann2006}.

Recently, a sizable fraction of thermalized muonium emitted into vacuum from mesoporous thin SiO$_{2}$ films has been reported \cite{Antognini2012}. The muonium vacuum yield per implanted $\mu^{+}$  with 5 keV implantation energy was measured to be 0.38(4) at 250 K and 0.20(4) at 100~K. The high muonium vacuum yield, even at low temperatures, is an important step towards new measurements with this atomic system and especially a more precise measurement of the muonium 1S-2S transition frequency \cite{Meyer2000}. In fact, such a source of cold muonium (100 K compared to 300 K of the previous measurements) opens the possibility of performing continuous wave (CW) laser spectroscopy of this transition because the larger vacuum yield compared to what has been previously reported \cite{MuW}-\cite{Janissen} and the increase in the interaction time between the atoms and the laser beam will compensate the lower power available compared to a pulsed laser. In fact, in the resonant weak field approximation the excitation probability is proportional to the interaction time squared.  CW spectroscopy would decrease the statistical and systematical uncertainties of the previous experiment~\cite{Meyer2000} since the broadening due to the laser chirp, the AC Stark effect and the residual first order Doppler shift related to pulsed laser spectroscopy would be eliminated. These effects resulted in a line width of about 20 MHz. 
Monte-Carlo simulations of the atoms' trajectories with numerical integration of the Bloch equations including photo-ionization and the AC Stark effect \cite{haas2006} show that the expected line width   for a CW experiment is at the level of 1 MHz, approaching the 145 kHz natural line width of the 1S-2S transition in muonium. The main source of broadening is due to the mean finite interaction time of the atoms with the laser beam which for the expected laser beam waist $\omega_0$=200 $\mu$m and a mean muonium velocity at 100 K of $v_{M}=2200$ m/s will be about 130 ns.
An enhancement cavity as in use at Max-Planck Institute of Quantum Optics in Garching for hydrogen excitation in the 2S state \cite{MPQ}, will be used to generate 4 W of laser power at 244 nm. The use of the cavity will grant a high degree of collinearity of the counter-propagating photons thus reducing the first order Doppler shift to a negligible level \cite{MPQ}. The broadening due to the AC Stark effect for the expected laser intensity $I$ can be estimated to be $\Delta \nu_{\text{AC}}=1.67 \cdot I $ Hz$\cdot$cm$^2/W  = 17$ kHz.

\begin{figure}[htbp]
\centering
\includegraphics[width=0.5\textwidth]{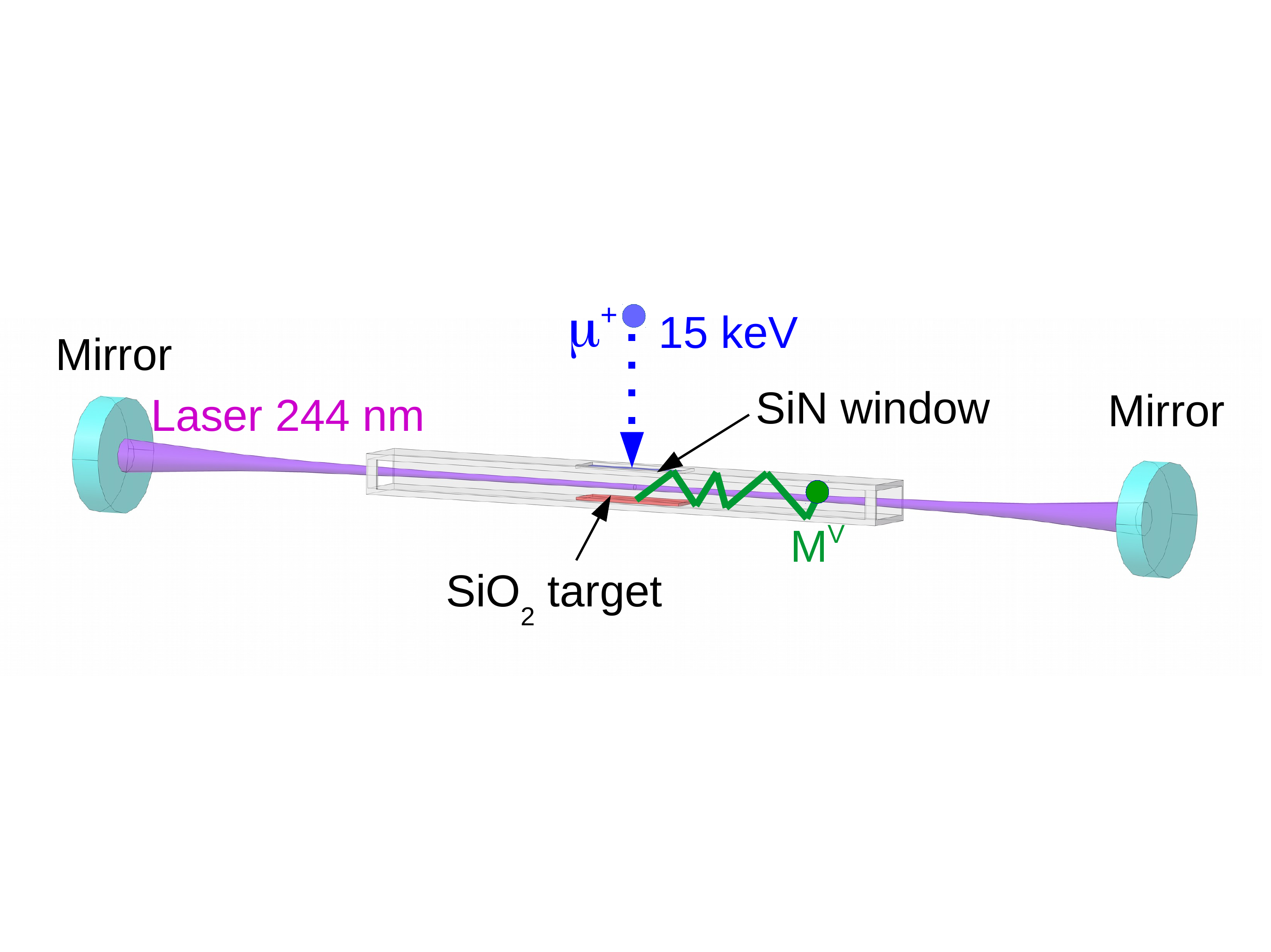}
\caption{Sketch of the muonium excitation in a confining volume. Muonium exiting the thin SiO$_2$ film into vacuum (M$^V$) is forced to bounce back and forth between the confining walls and thus multi-passages through the laser beam can occur enhancing its excitation probability.}
\label{MuExcScheme}
\end{figure}

 In this paper,  we demonstrate the feasibility of muonium confinement in a small volume/channel that allows one to perform such a measurement with currently available technology. 
Muonium confinement has the advantage of increasing the number of atoms crossing the laser beam if this is positioned along the axis of the channel (see Fig.~\ref{MuExcScheme}).  In this way the probability to laser excite a muonium atom is substantially enhanced leading to an improvement in signal rate.  A similar scheme has been recently employed for positronium (Ps, the electron-positron bound state) to detect Ps atoms excited in the 2S states \cite{hype15}. \\

With the Low Energy Muons (LEM) beam line at PSI and the $\mu^{+}$/M converter geometry we report here, an improvement  in precision of the 1S-2S transition frequency of more than an order of magnitude is in reach with a commercially available UV source \cite{toptica} combined with an enhancement cavity.\\

\section{Experimental technique and setup}
\begin{figure}[htbp]
\centering
\includegraphics[width=0.5\textwidth]{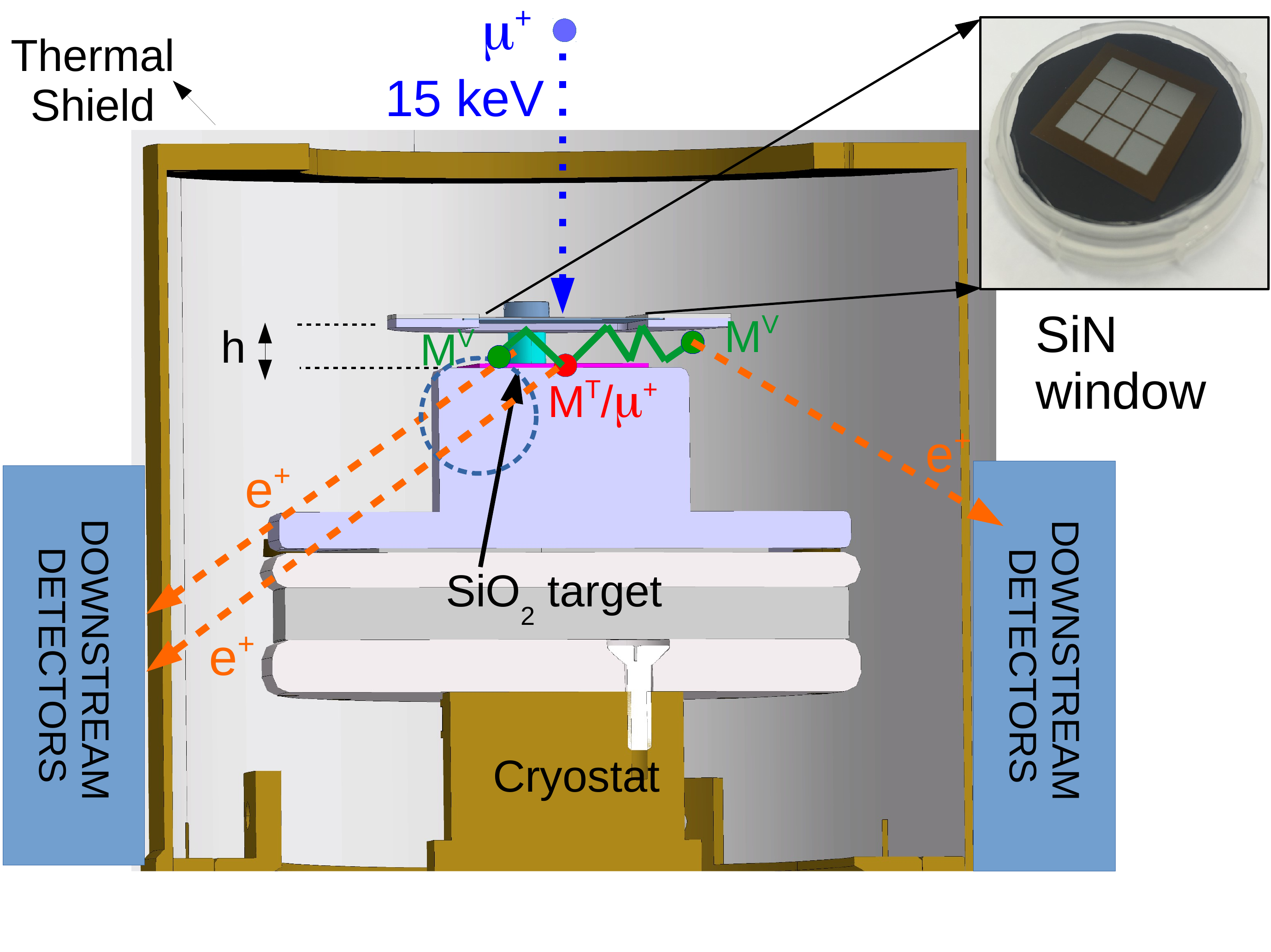}
\caption{Confinement of muonium between two surfaces. 
Sketch of the experimental setup and principle of the positron shielding technique (PST) technique. After crossing the thin SiN entrance window (shown in the inset) the 15 keV $\mu^{+}$ stop in the mesoporous SiO$_{2}$ thin film. 
Positrons from muons or muonium (M$^T$) decaying in the target have a constant detection probability. M exiting the target into vacuum (M$^V$) have an increased and time dependent detection probability because of the shielding/shadowing effect of the copper sample holder (shown as a dotted/blue circle).}
\label{fig:MuConfinement}
\end{figure}

The setup used to test the M confinement is sketched in Fig.~\ref{fig:MuConfinement}. It consists of two aluminum thin plates of 0.5~mm and 1.0~mm sandwiching a $3 \times 3$ array of 50 nm-thick $5.6 \times 5.6$ mm$^2$ SiN membranes (Silson, Blisworth, U.K.) that have 0.2~mm support ribs (see Fig.~\ref{fig:MuConfinement} (Inset)). To implant the keV muons in the confining cavity we used SiN windows that are quite robust (e.g. compared to carbon foils) and can be easily coated.
The SiN membrane has an overall area of $17.6 \times 17.6$ mm$^2$ and an average transmission of $> 80$\%. The membrane is installed in front of the mesoporous SiO$_{2}$ thin-film which is glued to the sample holder. Pillars with adjustable heights are used to vary the distance ($h$) between mesoporous silica and SiN window. The aluminum supports were coated with 3-4 nm of SiO$_2$ on the side where M is confined.
To avoid charging, the SiN window was coated with the same thickness of gold on the side of the incoming muons. The Cu sample holder was cooled down to 20 K repeatedly, and no mechanical damage due to thermal stress has been observed on the SiN membrane.  

The mesoporous sample used in this study is the same as the one we measured previously \cite{Antognini2012} for which the M vacuum yield $F_\mathrm{M}^\mathrm{vac}$  is 40\% at 250 K and 20\% at 100 K. As we have shown, the best fit to the data was achieved modelling M emission into vacuum with a cosine angular distribution and a Maxwell-Boltzmann kinetic energy according to the sample temperature. At T= 20 K, the M is decaying in the target because at this temperature it is adsorbed (sticks) to the SiO$_2$ surface and therefore does not diffuse into the vacuum  \cite{Antognini2012,Harshman1986a,Kiefl1984}. These parameters were used to simulate the M production at the mesoporous sample. 

\label{sec:level2}

\begin{figure}[htbp]
\centering
\includegraphics[width=0.49\textwidth]{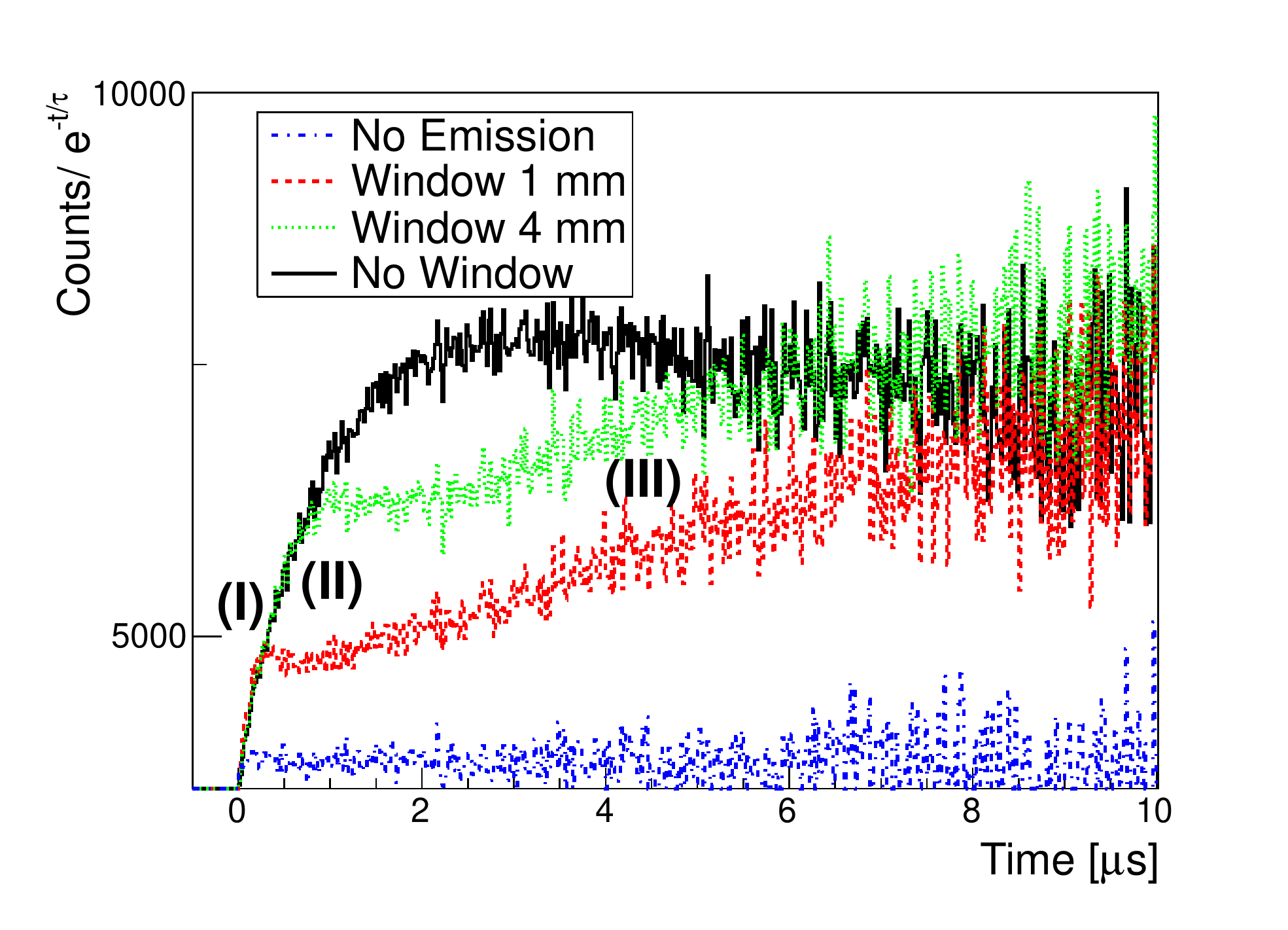}
\caption{Simulated time spectra of the downstream detectors after subtraction of the muon-decay for different cases: i) Dotted-dashed/blue line: No M emission into vacuum, ii) Dotted/red line: emission of Mu into vacuum at 250 K and cosine reflection at the walls with  $h$=1 mm separation between the SiN entrance window and the mesoporous SiO$_{2}$ thin film. iii) Green/dashed line: same as (ii) but with $h$=4 mm. iv) M Emission into vacuum at 250 K without SiN window. The three regions (I, II, III) defined on the graph are described in detail in the text.}\label{fig:SimMuConfinement}
\end{figure}

The principle of the measurements to test the muonium spatial confinement between two surfaces is sketched in Fig.~\ref{fig:MuConfinement}. A fraction of the low energy $\mu^{+}$ from the LEM beam line \cite{LEM} (5000 $\mu^{+}$/s with a beam size of 5 mm) are implanted through the thin (50 nm) SiN entrance window and stop in the mesoporous thin-film where about 60\% of them form muonium in the bulk material \cite{Antognini2012,Dehn:2014yra,Bakule:2013poa}. TrimSP simulations, validated with previous measurements with muons \cite{TrimSP}, predict that about 96\% of the $\mu^{+}$ with an initial energy of 15 keV are transmitted through the SiN window and implanted in the mesoporous silica ($\rho=1.1$ g/cm$^3$) with a mean energy of 4.9 keV (1.0 keV RMS) which corresponds to a mean implantation depth of about 80 nm (26 nm RMS). The formed muonium atoms diffuse through the interconnected porous network and a fraction of them reach the thin-film surface before the muon decays and are emitted into vacuum ($10^{-9}$ mbar) where muonium is free to move until it reaches the confining surface surrounding the target. The main goal of this work is to study the effect of the muonium hitting the walls in order to assess the feasibility of its confinement.  
 If reflections at the walls occur, the M atoms will diffuse from the center of the setup in the radial direction. Therefore with increasing time, they approach the plastic scintillators which surround the sample plate used to detect the positrons from the muon decays (the downstream detectors shown in Fig.~\ref{fig:MuConfinement}, see \cite{LEM} for more details).
The positrons emitted from muonium which traveled a larger radial distance from the target before decaying have a considerably higher probability to be detected in the downstream positron counter because of the reduced shielding by the copper sample plate as schematically illustrated in Fig.~\ref{fig:MuConfinement}. 
Therefore, if muonium diffuses the time spectra of the downstream detector are distorted as a consequence of the position-dependent detection probability as shown in Fig.~\ref{fig:SimMuConfinement}.  These are typical positron shielding technique (PST) time spectra simulated with Geant4 \cite{geant4}  that were observed in \cite{Antognini2012}. The geometry of the copper sample holder was optimized in order to enhance the contrast for M atoms emitted into vacuum compared to M or $\mu^+$ decaying in the sample (see Fig.~\ref{fig:MuConfinement}), i.e. compared to the standard sample holder used in the LEM for muon-spin rotation the sample position was moved by 19 mm downstream and the holder diameter was reduced by a factor 2 \cite{KimPhD}.

In order to better visualize the effect of muonium atoms emitted into vacuum and reflecting between the SiN window and the mesoporous SiO$_{2}$, the simulated positron time spectra are divided by $e^{-t/\tau_\mu}$ to eliminate the muon lifetime ($\tau_\mu$=2.2 $\mu$s) effect as shown in Fig.~\ref{fig:SimMuConfinement}. 

For $\mu^{+}$ or muonium stationary on the sample plate, the detection efficiency of the decay positron as seen from the downstream detector remains constant in time. Hence a straight horizontal line is expected as shown in Fig.~\ref{fig:SimMuConfinement} (dashed-dotted/blue histogram). For M emission into vacuum without SiN entrance window, an increase in the spectra is expected (solid/black histogram) because of the smaller shielding effect of the sample holder for muonium exiting the target. 

\begin{figure}[htbp]
\centering
\includegraphics[width=0.49\textwidth]{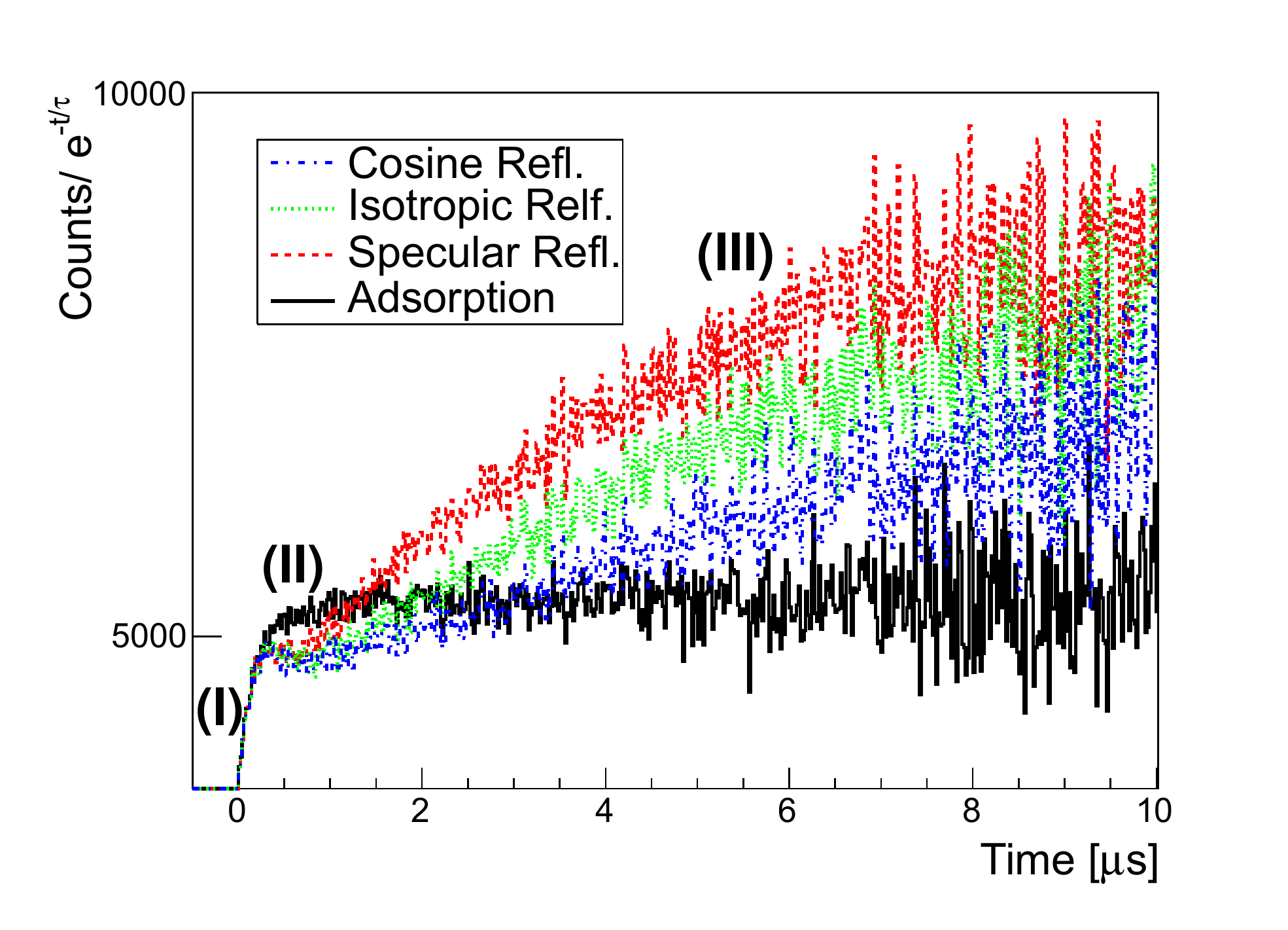}\hfill
\caption{
Simulated time spectra for different reflection scenarios of muonium emitted into vacuum at 250 K: no reflection, i.e. adsorption/sticking of muonium on the confining walls (black/solid curve), specular (red/dashed curve), isotropic (green/dot curve) and cosine distributed relative to the surface normal (blue/dot-dashed curve) reflections.}
\label{fig:SimMuReflection}
\end{figure}

When a SiN window is installed, the time evolution is more complicated. After emission into vacuum, the M atoms propagate freely until they reach the SiN membrane. Therefore, the time spectra in the first few hundred ns  (in region (I) in Fig.~\ref{fig:SimMuConfinement}) are "identical" to the histograms computed  without the window. 
When they reach the window, the increase in the detection efficiency stops because the muonium are either adsorbed at the wall or back-reflected. If adsorption occurs, a flat curve is obtained (see black/solid curve in Fig.~\ref{fig:SimMuReflection}). 
If back-reflection occurs this result first in a sudden drop in positron counts as visible in region (II) of Figs.~\ref{fig:SimMuConfinement} and \ref{fig:SimMuReflection} because the muonium atoms move back to the sample holder plate where the shielding for the decaying positrons is higher. With time the muonium atoms will radially drift towards the surrounding positron detectors causing the increase in the time spectra at later times (region (III)). 

From the increase related with this slow diffusion one can derive information on the reflection process and the adsorption probability. To illustrate the sensitivity to the reflection model different reflection scenarios of the muonium atoms at the confining walls, i.e. specular, isotropic and reflections distributed according to a cosine distribution were simulated (see Fig.~ \ref{fig:SimMuReflection}).

\section{Results}
\label{sec:level4}

\begin{figure*}[htbp]
\begin{center}
\includegraphics[width=0.45\textwidth]{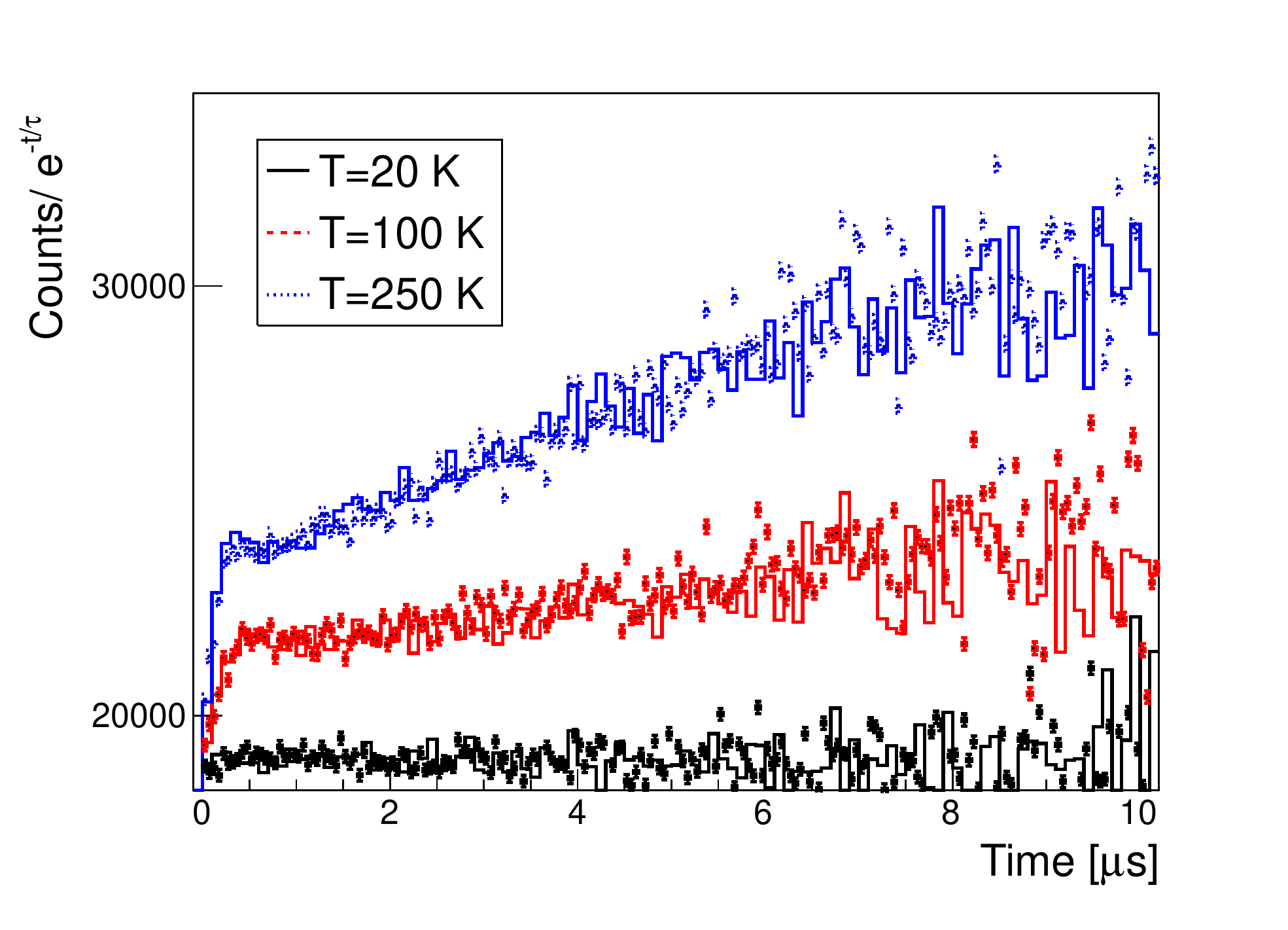}
\includegraphics[width=0.45\textwidth]{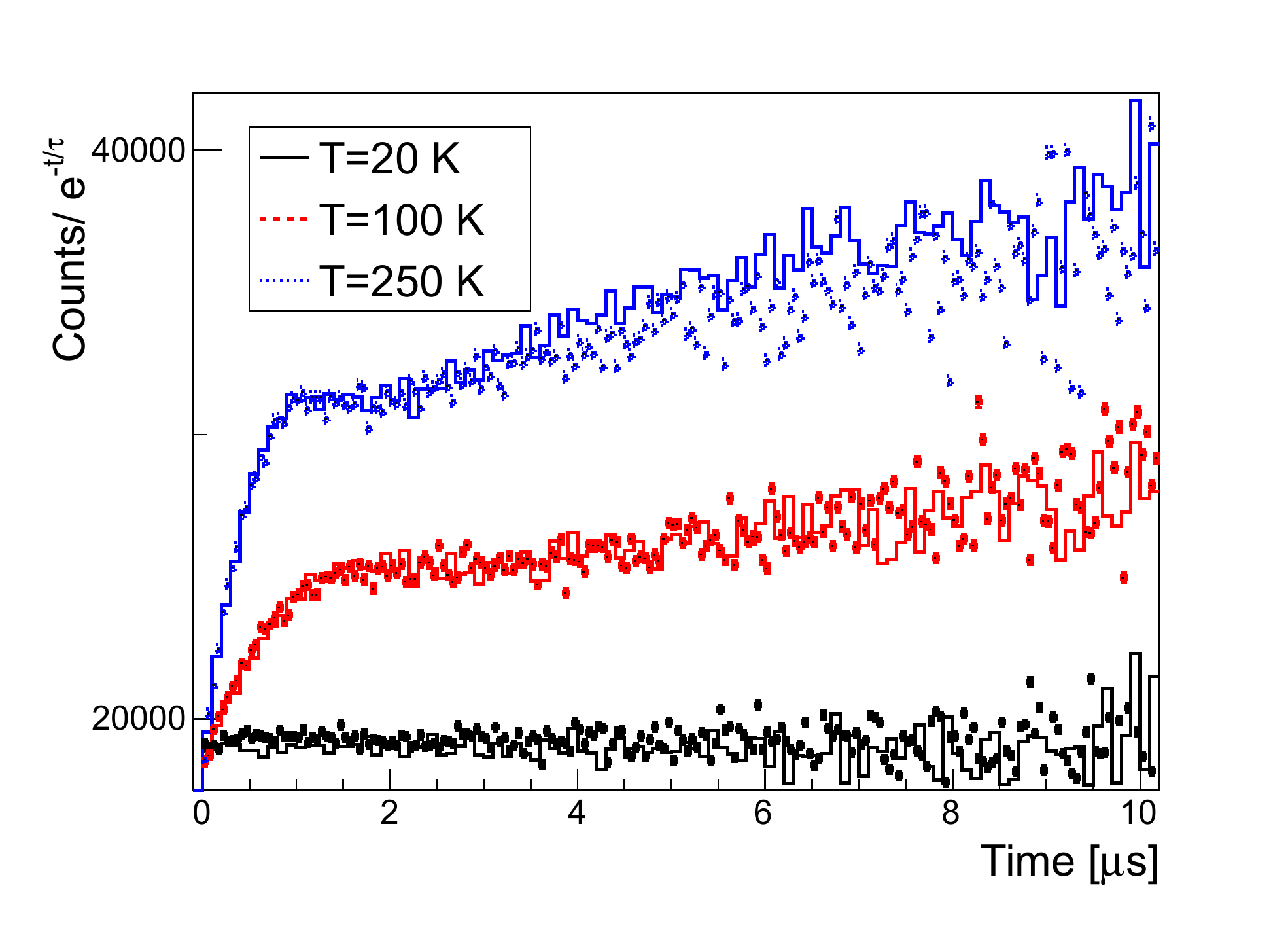}

\caption{The points are the measured time spectra at $T=20$, $100$ and $250$~K for a distance between the SiO$_2$ target and the SiN window of  $h=$1~mm (left) and of $h=$4~mm  (right). The solid lines are the corresponding simulated time spectra (see text for more details).}
\label{fig:plot2014-data}
\end{center}
\end{figure*}

The positron time spectra of the downstream detectors for measurements done at $h=1$~mm and $h=4$~mm separation and various temperatures are shown in Fig.~\ref{fig:plot2014-data}. The solid line represents the simulated time spectra which result in a good agreement with the data. As expected from our previous results \cite{Antognini2012}, the yield of muonium emitted into vacuum increases with temperature, i.e. in the simulation the $F_\mathrm{M}^\mathrm{vac}$ is 20\% at 100 K and 40\% at 250 K. For times $t \in [0;0.5]$~$\mu$s the time spectra increase quickly until a ``plateau'' is reached. This ``plateau'' level is proportional to the muonium emission probability into vacuum $F_\mathrm{M}^\mathrm{vac}$, whereas the linear coefficient of the slope at early times depends on the initial velocity of the muonium atoms and therefore on the sample temperature \cite{Antognini2012}.  
 The time when the muonium hit the SiN window is easily visible in the time spectra and increases, as one would expect, with the distance between the mesoporous silica and the entrance window. This is about 0.25 $\mu$s for $h=1$ mm and 1 $\mu$s for  $h=4$ mm. 
 For later times $t \in [1;10]$~$\mu$s, the histograms show a slow increase. This demonstrates clearly that the muonium atoms can radially diffuse inside the confinement region and survive the collisions with the walls. This is nicely evidenced by the comparison with the 20~K data shown in Fig.~\ref{fig:plot2014-data} where M is produced but it does not diffuse out of the sample \cite{Antognini2012} because it sticks to the SiO$_2$ surface. Hence, the data down to a temperature of 100~K clearly indicate M bouncing between the two walls and thus the feasibility of its confinement in a small volume.

\begin{figure*}[htbp]
\begin{center}
\includegraphics[width=0.45\textwidth]{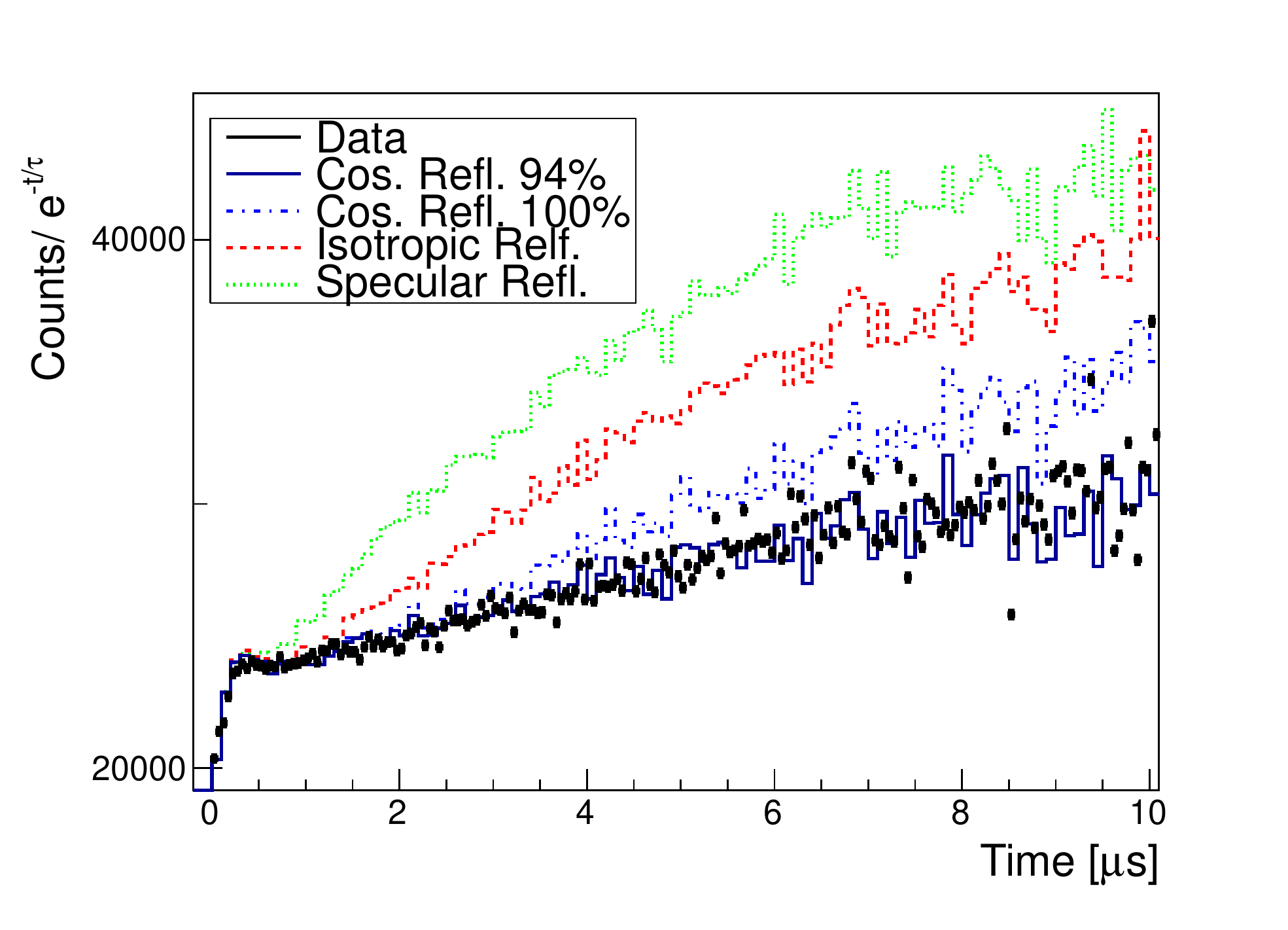}
\caption{The points are the data at  $250$~K for a distance between the SiO$_2$ target and the SiN window of  $h$=1~mm. The lines are the simulated time spectra for different reflection scenarios  (see text for more details).}
\label{fig:data_refl}
\end{center}
\end{figure*}

The increase at later times does not depend only on the muonium kinetic energy distribution but also on the type of reflection and on the reflection probability. Thus the slope at delayed times contains information on the reflection process and its probability $R$. 
The best fit to the data are obtained simulating a reflection following a cosine distribution (see Fig.~ \ref{fig:data_refl}) as predicted by kinetic theory of gases \cite{knudsen} and supported by recent advanced simulations of atoms reflection at rough surfaces \cite{cosine}. Assuming this is the correct distribution, a probability of $R_{250K}=94\pm4$\% at 250 K and $R_{100K}=90\pm6$\% at 100 K is extracted. These reflection probabilities are compatible with the fraction of atoms with kinetic energies below about 4-5 meV for muonium emitted with Maxwell Boltzmann distributions at 100 K and 250 K, respectively.  This energy corresponds to about 50 K, a substrate temperature at which muonium is known to stick to SiO$_2$. This effect was first observed in silica powder by measuring the hyperfine splitting (HFS) as a function of the temperature. It was noticed that below 100 K the value of the HFS was dropping from the vacuum value and it was shown that this was consistent with thermal adsorption at the SiO$_2$ \cite{Kiefl1984}. In mesoporous silica films, an abrupt drop of the M emitted into vacuum below 100 K has been observed. Because the muon spin rotation data show a constant muonium formation probability as a function of the temperature, this was also interpreted as adsorption of M atoms on the surface of the pores \cite{Antognini2012}. 
In light of those results, the efficient reflection of M with energies of 4-5 meV on SiO$_2$ surfaces  could be anticipated.
 A possible explanation for the observed reflection probabilities (summarised in Table \ref{tab1}) is that muonium atoms with kinetic energies smaller than the depth of the Lennard-Jones potential (which can be estimated with our previous results) are adsorbed at the SiO$_2$ walls also at the respective higher substrate temperatures. 

\begin{table}
\caption{Summary of number of the reflection probabilities at the surface and the production of confined M atoms (1/s) using the LEM beamline and the target geometry of Fig. \ref{fig:MuConfinement} for different temperatures.}
\begin{ruledtabular}
\begin{tabular}{ccc}
Temperature (K) & M (1/s) & Refl. Prob.(\%) \\
\hline
20 & 0 & 0 \\
100 & 768$\pm$ 31 &90$\pm$6  \\
250   & 1459$\pm$ 58 & 94$\pm$4 \\
\end{tabular}
\end{ruledtabular}
\label{tab1}
\end{table}


\section{Summary and conclusions}
\label{sec:level5}
The comparison of the data and the simulation shows that the reflection effect of muonium between the confining surfaces is very significant compared to the adsorption effect and follows a cosine distribution. 
The measured time spectra show that muonium atoms are reflected from confining surfaces even at 100~K.  
At this temperature and for a distance of $h=$1 mm, this is estimated to give an enhancement of a factor 5 in the excitation probability for the muonium 1S-2S transition compared to a geometry in which the atoms would just pass only once through the laser beam. Therefore, the possibility of muonium confinement which has been demonstrated here opens the way to a measurement of the muonium 1S-2S transition frequency with existing technology at the 0.2 ppb level (an improvement of a factor 20 compared to \cite{Meyer2000}). This will provide a test of bound-state QED (the theoretical uncertainty is 0.4 ppb \cite{pachucki,karshenboim1S2S}), the best verification of charge equality in the first two generations of particles and an improved determination of the muon mass at the 40 ppb level (a factor 3 better than currently determined \cite{LiuHFS}). Such a measurement would be limited by the statistical uncertainty and therefore an even higher accuracy could be expected in the near future because of the ongoing efforts to develop high flux and brightness slow muon beams \cite{MuCool,Strasser:2014dsa}.

\section*{Acknowledgments}
This work has been supported by the Swiss National
Science Foundation under the grant numbers 200020\_166286 and 200020\_159754 and by the ETH Research grant ETH-35 14-1.
We also acknowledge the help of the PSI and ETH Zurich IPP workshops and support
groups. Special thanks to A. Gendotti, M. Horisberger and A. Suter. The LEM measurements have been performed at the Swiss Muon Source S$\mu$S, PSI.





\end{document}